\documentclass[aps,prl,twocolumn]{revtex4}

\usepackage{graphicx}
\usepackage[hang]{subfigure}
\usepackage[utf8]{inputenc}

\begin{document}

\title{Tuning the fragility of a glassforming liquid by curving space}

\author{Fran\c{c}ois Sausset}
\author{Gilles Tarjus}
\author{Pascal Viot}

\affiliation{Laboratoire de Physique Th\'eorique de la Mati\`ere Condensée, Université Pierre et Marie Curie-Paris 6, UMR CNRS 7600, 4 place Jussieu, 75252 Paris Cedex 05, France}

\begin{abstract}
We investigate the influence of space curvature, and of the associated ``frustration'', on the dynamics of a model glassformer: a monatomic liquid on the hyperbolic plane. We find that the system's fragility, \textit{i.e.} the sensitivity of the relaxation time to temperature changes, increases as one decreases the frustration. As a result, curving space provides a way to tune fragility and make it as large as wanted. We also show that the nature of the emerging ``dynamic heterogeneities'', another distinctive feature of slowly relaxing systems, is directly connected to the presence of frustration-induced topological defects.
\end{abstract}

\maketitle

Among the many anomalous properties associated with glass formation, ``fragility'' is one that has attracted much attention \cite{Angell:1995,Debenedetti:2001,Martinez:2001,Xia:2000,Sastry:2001,Garrahan:2003,Tarjus:2005}. Large fragility, \textit{i.e.} large deviation of the temperature dependence of the viscosity and of the structural relaxation time from an Arrhenius behavior, is usually taken as the signature of a collective phenomenon that grows as temperature decreases. This is certainly one incentive for the continuing search for a theory of the glass transition \cite{Debenedetti:2001,Xia:2000,Sastry:2001,Garrahan:2003,Tarjus:2005}. Yet, the absence of a simple glassforming liquid model in which one can control the degree of fragility, hence the extent to which collective behavior develops, has hindered progress on developing and testing candidate theories.

Since the early work of Frank\cite{Frank:1952}, a promising line of research on supercooled liquids and the glass transition has relied on the concept of ``geometric frustration''\cite{Sadoc:1999,Nelson:2002,Tarjus:2005}. Frustration in this context can be defined as an incompatibility between extension of the local order preferred in a liquid and tiling of the whole space. The paradigm is the icosahedral order in metallic liquids and glasses, which although locally favored cannot tile space due to topological reasons\cite{Frank:1952}. Frustration of the icosahedral order, however, can be suppressed by leaving the Euclidean world and curving space\cite{Sadoc:1999,Nelson:2002}. In a series of insightful articles\cite{Nelson:2002,Nelson:1983a,Rubinstein:1983}, Nelson and collaborators have proposed a simpler two-dimensional ($2D$) analog: by placing a liquid of disks on a $2D$ manifold of constant negative curvature (the hyperbolic plane), the local hexagonal order that can tile the ordinary Euclidean plane is now frustrated in a way which mimics by many aspects the frustration of icosahedral order in $3D$ Euclidean space. The model of a monatomic liquid on the hyperbolic plane therefore offers the opportunity to investigate, at a microscopic level, the influence of the degree of frustration, here controlled by the curvature, on the slowing down of the relaxation associated with glass formation.

\begin{figure*}[t]
\subfigure[\hspace{3.5cm}]{\label{fig:Poincare}\includegraphics[width=4cm, trim= 0cm 1.3cm 0 0]{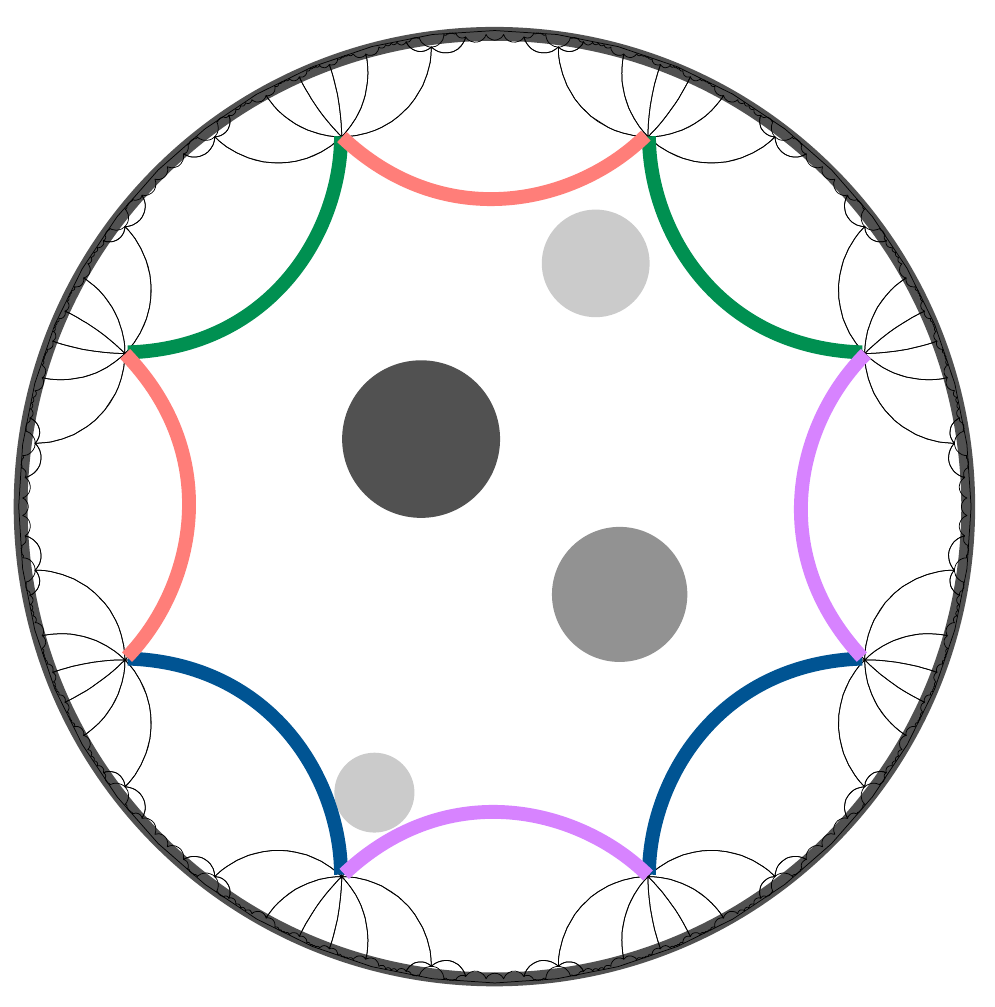}}
\hspace{2cm}
\subfigure[\hspace{9cm}]{\label{fig:Fs}\includegraphics[width=9cm, trim= -1.5cm 3.2cm 1.4cm 0.2cm]{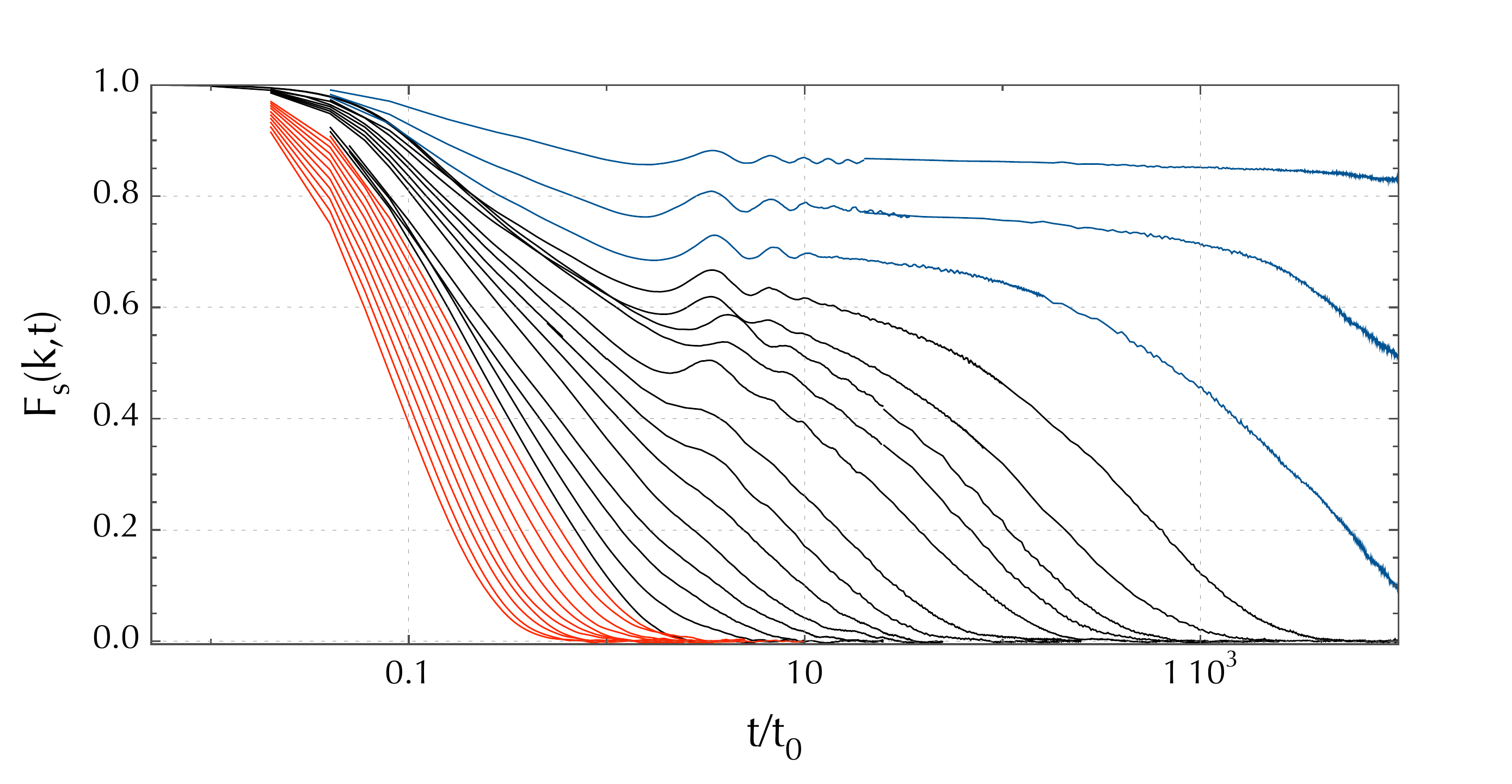}}
\caption{(a) The projection of $H^2$ associated with the Poincar\'e disk model: atoms are represented by disks whose diameter contracts as they move away from the center of the Poincar\'e disk. The geodesics are arcs of circles perpendicular to the disk boundary. We also display the octagonal tiling (in which $8$ octagons meet at each vertex) used in the pbc. (b) $F_s(k,t)$ versus $t$ for $\kappa\sigma=0.2$ and for different temperatures $T/T^*$ ranging from $3.9$ to $0.1$ (left to right): for $T\sim T^*$ and below,  stretched exponential behavior, $\exp(-(t/\tau)^\beta)$, is observed with $\beta$ decreasing from $1$ at the highest $T$ to $0.54$ at the lowest equilibrated $T$; for $T<0.35\,T^*$ the system is out of equilibrium and ages (last three curves).}
\end{figure*}

We present the results of the first computer simulation of the dynamics of a liquid in curved hyperbolic space. The hyperbolic plane $H^2$, also called pseudo-sphere or Bolyai-Lobatchevski plane, is a Riemannian surface of constant negative curvature\cite{Hilbert:1952,Coxeter:1969}. Contrary to a sphere, which is a surface of constant positive curvature, $H^2$ is infinite: this allows one to envisage the thermodynamic limit at constant curvature. However, $H^2$ cannot be embedded as a whole in the $3D$ Euclidean space and ``models'', \textit{i.e.} projections, must be used for its vizualization. The hyperbolic metric is often given in polar coordinates $(r, \phi)$, namely,
\begin{equation}
\mathrm{d}s^2= \mathrm{d}r^2 + \left(\frac{\sinh(\kappa r)}{\kappa} \right)^2 \mathrm{d}\phi^2,
\end{equation} 
which makes apparent the connection with the more familiar metric of the sphere $S^2$ that is obtained by replacing the parameter $\kappa$ by $i \kappa$. The Gaussian curvature of $H^2$ is $-\kappa^2<0$; $\kappa$ therefore measures the deviation from flat space and $\kappa^{-1}$ can be taken as an intrinsic frustration length. We consider the Poincar\'e disk model (Fig. \ref{fig:Poincare}) that maps the whole infinite space $H^2$ onto the open disk of radius unity. This projection ($r'=\tanh(\kappa r/2), \phi'=\phi$) is conformal, \textit{i.e.} it preserves the angles, but is not isometric: the Euclidean distance between two points of the disk separated by a given distance in $H^2$ shrinks to zero when the points approach the disk perimeter.

To carry out a Molecular Dynamics (MD) simulation of particles on the hyperbolic plane, a number of serious methodological problems have to be resolved, which we only briefly allude to. Once the model is properly defined, the two main ingredients in any MD simulation are the algorithm to solve the Newton equations of motion and the boundary conditions, usually chosen as periodic in order to more rapidly converge to the thermodynamic limit corresponding to the experimental situation. Among the peculiarities one encounters when leaving flat space to consider curved manifolds such as $H^2$ is the absence of a global definition of parallel vectors. We handle this and generalize the standard MD algorithm to the hyperbolic plane by using a method detailed in a forthcoming publication. Even more delicate is the question of the periodic boundary conditions (pbc's). Due to the hyperbolic nature of the metric, the contribution of the boundary of any finite system is always of the same order of magnitude as that of the bulk of the system. Implementing proper pbc's is therefore crucial. Again, one must account for the specificities of hyperbolic space: first, an infinite number of regular tilings of $H^2$ are possible, and second, the area of the elementary cell of a given tiling is fixed by the curvature (see below)\cite{Hilbert:1952,Coxeter:1969}. As a consequence of the latter property, studying finite-size effects at constant curvature requires to change the boundary condition. Building on our earlier work\cite{Sausset:2007}, we have implemented two different pbc's: an octagonal pbc (Fig. \ref{fig:Poincare}) and a pbc with a larger unit cell formed by a regular $14$-gon with a specific pairing of the edges.

We have studied a monatomic liquid model in which atoms pairwise interact via the standard truncated Lennard-Jones potential $v(r) = 4\epsilon\left( (\sigma/r)^{12}-(\sigma/r)^{6} \right) $ (with a cut-off at $2.5 \sigma$), where the distance $r$ is defined with the hyperbolic metric. The control parameters are the temperature $T$, the density $\rho$, and the frustration associated with space curvature and characterized by the dimensionless parameter $\kappa \sigma$. As already stressed, there are a number of geometrical and topological constraints associated with the pbc. For instance, the area of the unit cell is $A=4 \pi \kappa ^{-2} (g-1)$, where $g$ is the genus of the compact manifold associated with the pbc and is equal to $2$ for the octagon and $3$ for the $14$-gon. For a given density, the number $N$ of atoms is thus fixed by the curvature and the pbc: in the present work, $N$ typically varies between $300$ and $30\, 000$ as one decreases the frustration. We are interested by weak frustration for which the local order remains hexagonal as in the Euclidean plane. Indeed, for large enough frustration, the preferred arrangement of atoms around a central one is no longer a hexagon, but a heptagon, with unfrustrated extension of the heptagonal order to the whole space; on further increasing the frustration, one encounters locally preferred arrangements formed by polygons with an increasingly larger number of sides \cite{Rubinstein:1983}. (Disk packings for large negative curvature have also been recently considered in \cite{Modes:2007}.) Simulations are performed for $\kappa \sigma$ spanning one order of magnitude from $0.02$ to $0.2$. In addition to computing usual static quantities, \textit{e.g.} the pair correlation function, we have carried out a direct analysis of the topological defects, which will be discussed below, and we have monitored several dynamic observables characterizing the motion of the atoms. From the distance travelled by any atom $j$ between two times $t'$ and $t'+t$, $d_j(t',t'+t)$, we compute the hyperbolic generalization of the self intermediate scattering function,
\begin{equation}
F_s(k,t) = \frac{1}{N}\sum_{j=1}^{N}\langle P_{-\frac{1}{2}+i\frac{k}{\kappa}}(\cosh(\kappa d_j(0,t)))\rangle,
\end{equation}
where $P_{-\frac{1}{2}+i\frac{k}{\kappa}}$ is a Legendre function of first kind (such that Eq. (2) reduces to the conventional spatial Fourier transform in the Euclidean limit\cite{Terras:1985}).

\begin{figure}[b]
\includegraphics[width=7cm, trim= 0 0.8cm 0 0.8cm]{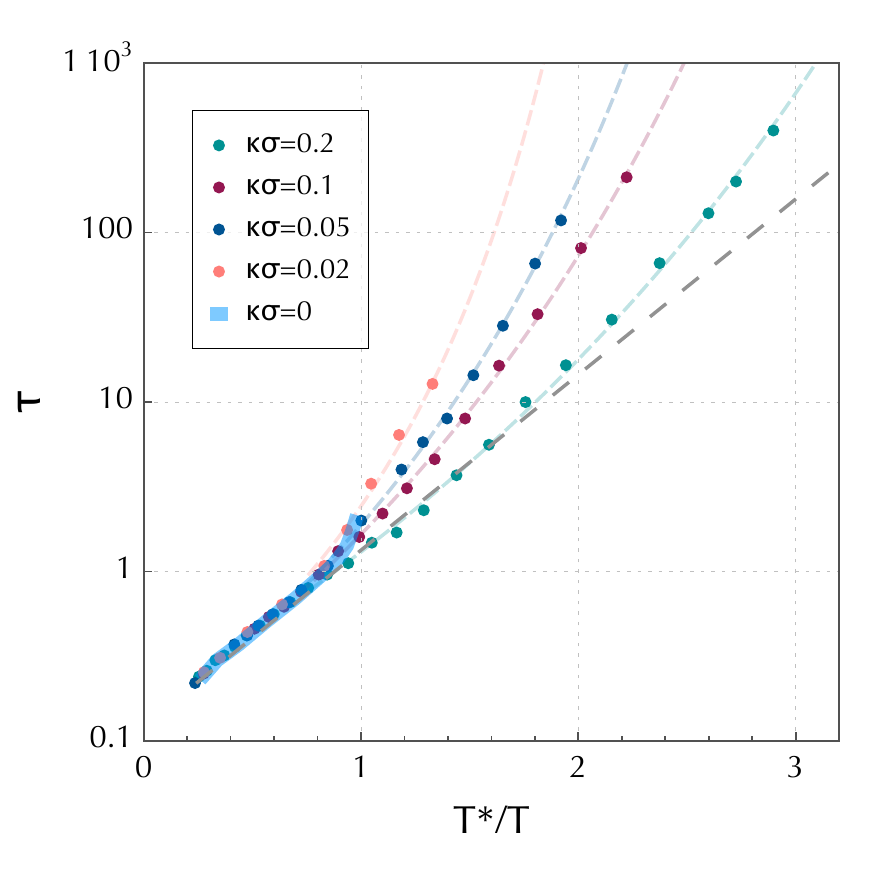}
\caption{\label{fig:fragility}Arrhenius plot of the translational relaxation time $\tau$ versus $T^*/T$ for several curvatures, where $T^*$ is the (approximate) location of the ordering transition in the Euclidean plane. The particle density is $\rho \sigma^2=0.85$. Note the deviation from the simple Arrhenius dependence shown by the dashed line. As the frustration $\kappa \sigma$ decreases, the deviation is stronger, \textit{i.e.} the fragility gets larger. The thick blue line above $T^*$ corresponds to the Euclidean case. The continuous lines are fits to the VFT formula, $\tau=\tau_0 \exp(D T_0/(T-T_0))$: as frustration decreases, $D$ is found to decrease from $10.8$ to $4.6$ while $T_0/T^*$ increases from $0.12$ to $0.33$, which again indicates an increase in fragility.}
\end{figure}

As a benchmark, we have first considered the Euclidean case ($\kappa=0$). We find that, irrespective of the cooling rate, the liquid orders in an hexagonal structure at a temperature $T^*(\rho \sigma^2)$ (\textit{e.g.} $T^*(\rho \sigma^2 = 0.85) \simeq 0.75$ in Lennard-Jones units). No glass formation is therefore possible, as anticipated. A very different behavior is observed when frustration is switched on by curving space. No ordering transition takes place (the transition at $T^*$ is thus ``avoided''\cite{Tarjus:2005}) and the liquid phase can now be cooled in equilibrium below $T^*$.

From $F_s(k,t)$ with $k$ chosen near the maximum of the static structure factor ($k\simeq  \sigma^{-1}$), we have extracted the translational relaxation time $\tau$, which is determined when $F_s(k,t)=0.1$ (see Fig. \ref{fig:Fs}). An alternative definition of $\tau$ is obtained from the fit of $F_s(k,t)$ (beyond the plateau) to a stretched exponential, $e^{-(t/\tau)^\beta}$. Up to a multiplicative constant, the two definitions of $\tau$ give similar results for the $T$ dependence. At low enough  $T$, one reaches the limit  of the computer resources and the liquid falls out of equilibrium to freeze in an amorphous solid, \textit{i.e.} a glass.

As shown in Fig. \ref{fig:fragility}, one observes a striking pattern of variation of fragility with frustration. At high $T$ above $T^*$, the data show no dependence on curvature. This is easily understood by combining the fact that the relaxation remains a local phenomenon and that locally $H^2$ appears flat for the atoms (recall that $\kappa\sigma\ll1$). A marked deviation from Arrhenius dependence, \textit{i.e.} a super-Arrhenius behavior, is found below $T^*$. The magnitude of this deviation unambiguously increases as the frustration parameter is reduced and one gets closer to the unfrustrated Euclidean case. As predicted by the frustration-limited domain theory\cite{Tarjus:2005}, fragility therefore increases as frustration decreases. A rationale for this trend is that as the system gets closer to the avoided transition, the spatial correlations associated with frustrated ordering grow larger; collective behavior thus occurs on longer length scales, which results in a more strongly super-Arrhenius dependence of the relaxation time and a larger fragility. Note that in line with this fragility pattern, the stretching exponent $\beta$ is found to decrease with decreasing frustration: at $T/T^*\simeq 0.85$, $\beta=0.64, 0.54,0.50,0.42$ for $\kappa\sigma = 0.2,0.1,0.05,0.02$.

A crude heuristic argument  suggests that the increase of fragility goes logarithmically with the inverse of the frustration. The idea is to compare the energy scales involved in the activation barriers for relaxation at low and high $T$ and derive an estimate of the fragility through their ratio. At high $T$, the scale is provided by the interaction energy between atoms and is independent of curvature as seen from Fig. \ref{fig:fragility}. On the other hand, one expects that the low-$T$ dynamical behavior is controlled by the motion of the rare frustration-induced defects (see below). An estimate for the associated energy scale is obtained by considering the continuum approach valid at sufficiently large wavelength and low $T$. The energy of frustration-induced disclinations (in an otherwise hexatic medium) is found to be a constant plus a term proportional to $\ln(1/\tanh(\kappa \sigma/2))$\cite{Nelson:2002}, which for small frustration behaves as  $\ln(1/(\kappa \sigma))$. The ratio of the energy scales at low and high $T$ therefore goes as the logarithm of $1/(\kappa \sigma)$, which is compatible with the variation of fragility obtained from the simulation data. This indicates that fragility can be made as large as wanted by taking the limit of vanishingly small curvature.

\begin{figure}
	\subfigure[\hspace{3.5cm}]{\includegraphics[width=4.2cm,trim=0.1cm 0.6cm 0.34cm 0]{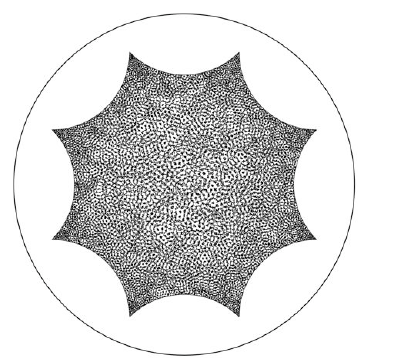}}
	\subfigure[\hspace{3.5cm}]{\includegraphics[width=4.2cm,trim=0.1cm 0.6cm 0.34cm 0]{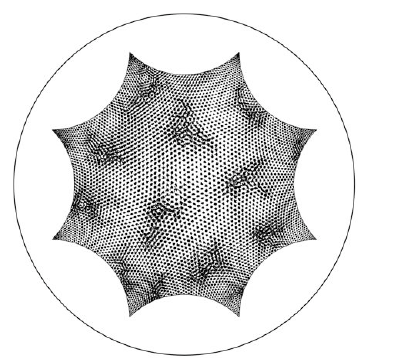}}
	\subfigure[\hspace{3.5cm}]{\includegraphics[width=4.2cm,trim=0.4cm 2.4cm 1.36cm 0]{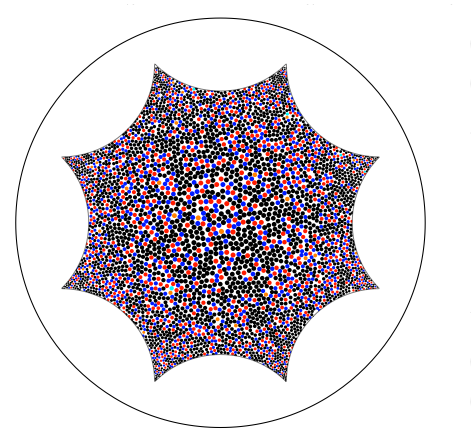}}
	\subfigure[\hspace{3.5cm}]{\includegraphics[width=4.2cm,trim=0.4cm 2.4cm 1.36cm 0]{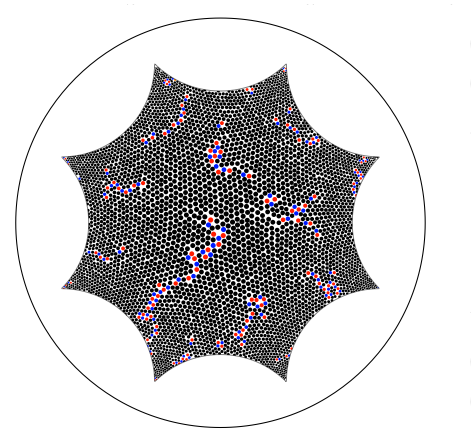}}
\caption{(a),(b) Atomic trajectories followed for a time interval during which the average distance travelled by the atoms is roughly $\sigma$ at (a) $T/T^*=2.4$ and (b) $T/T^*=0.52$. Whereas at high $T$ all atoms seem to move by a comparable amount and the dynamics is thus spatially homogeneous, a strikingly different picture is obtained at low $T$: most atoms hardly move or just rattle in the cage formed by their neighbors and mobility is concentrated in rare localized regions, illustrating the spatial heterogeneity of the dynamics. (c),(d) Same atomic configurations as in (a) and (b) (respectively), with the color indicating the coordination number for each atom. Black: 6 neighbors (local hexagonal order), red: 7 neighbors (disclination of topological charge $-\pi/3$ ), blue: 5 neighbors (disclination of charge $+\pi/3$); at $T/T^*=2.4$ (c), there are also defects with larger charges (orange: 8 neighbors, cyan: 4 neighbors). Note the correspondence at low $T$ between the rare localized defective regions in (d) and the regions of high mobility in (b). The frustration is $\kappa \sigma=0.05$ and the particle density $\rho \sigma^2=0.85$.}
\end{figure}

Another canonical feature of slowly relaxing systems, glassforming liquids in particular, is the ``heterogeneous'' nature of the dynamics\cite{Ediger:2000,Hurley:1995,Weeks:2000, Berthier:2005}.
This phenomenon is easily detected by following the particle trajectories for a given period of time, as shown in Figs. 3 (a),(b). Topological defects\cite{Reichhardt:2003} and/or medium-range ordering\cite{Shintani:2006} have been suggested as playing a role in disordered $2D$ phases exhibiting dynamic heterogeneities, and the present $2D$ monodisperse glassforming liquid offers a unique opportunity to investigate this point.

The topological defects can be defined at a microscopic level by analyzing the local environment of each atom. To do so, we use a curved-space generalization of the Voronoi tessellation that provides an unambiguous means to assign a number of nearest neighbors to each atom. Most atoms have $6$ neighbors, which corresponds to the hexagonal local order. Defects, more specifically point ``disclinations'', are then associated with atoms with a coordination number different from $6$. Negative curvature forces in an irreducible number of disclinations of negative topological charge (more than $6$ neighbors), this number being fixed by topological relations. In Figs. 3 (c),(d), we display the same atomic configurations as those plotted in Figs. 3 (a),(b) with a color code indicating the coordination number of each atom.  At high $T$, there is a large density of defects and it is hardly possible to notice the imbalance in favor of negative disclinations. At low $T$ on the other hand, the number of defects is small and one can clearly see the emergence of large domains of local $6$-fold order coexisting with small localized defective regions. A closer inspection reveals that there are exactly $12$ such regions consisting of a $7$-fold disclination and attached short strings of little dipoles of $5$-fold and $7$-fold disclinations forming ``dislocations''. Such strings have been dubbed ``grain boundary scars'' in the context of crystals on spherical surfaces\cite{Bausch:2003}. (Note that the irreducible \textit{number} of $7$-fold disclinations is fixed by the pbc, here $12(g-1)$ with $g=2$, but that the irreducible \textit{density} of disclinations decreases, and the typical size of the locally ordered domains increases, as curvature decreases.) Comparison of the two sets of figures clearly shows that the emerging heterogeneous character of the dynamics is directly linked to the topological defects: at low $T$, the regions of high mobility coincide with the vicinity of the intrinsic frustration-induced defects and their attached strings of dislocations whereas the regions of low mobility coincide with the hexagonal patches. We stress that the system is in a liquid phase even at low $T$ and that all defects and atoms move over long enough time spans. Preliminary results on the $4$-point space-time correlation function $\chi_4(t)$ obtained as the variance of the local relaxation associated with $F_S(k,t)$\cite{Dasgupta:1991} indicate that the spatial correlations in the dynamics are maximum around a timescale of the order of the relaxation time $\tau$ and increase continuously as $T$ decreases. The phenomenology is thus similar to that found in other glassformers\cite{Ediger:2000,Hurley:1995,Weeks:2000,Berthier:2005}.

The above results suggest the passage as $T$ decreases from a local atomic dynamics to a collective relaxation controlled by the motion of topological defects, with an intermediate region that is determined by the proximity to the avoided transition at $T^*$. This intermediate region becomes more important as frustration decreases and one expects that in this regime, growth of static spatial correlations, super-Arrhenius behavior and extension of the dynamic heterogeneities all go together. However, the extent of static spatial correlation saturates to a value given by the intrinsic frustration length $\kappa^{-1}$ (which fixes the average distance between the remaining intrinsic disclinations); at low $T$ and over distances beyond $\kappa^{-1}$, the slowing down of relaxation should therefore be controlled by the rare intrinsic $7$-fold disclinations. One may speculate that in this regime the growing dynamic correlations (as extracted from $\chi_4(t)$) decouple from the static spatial correlations and reflect longer-range correlation among the mobilty regions associated with the residual defects. Work is now in progress to investigate this potential low-$T$ decoupling phenomenon.

\end{document}